\documentstyle[12pt]{article}
\renewcommand{\baselinestretch}{1.2}
\begin{document}
%
%
%
\pagenumbering{arabic}
\setcounter{page}{1}
\centerline{\large \bf A Comparative Study of}
\centerline{\large \bf Some Pseudorandom Number Generators}

\vskip1cm
\centerline{I. Vattulainen$^1$, K. Kankaala$^{1,2}$, J. Saarinen$^1$,
 and T. Ala-Nissila$^{1,3}$}
\bigskip
\centerline{\em $^1$Department of Electrical Engineering}
\centerline{\em Tampere University of Technology}
\centerline{\em P.O. Box 692}
\centerline{\em FIN - 33101 Tampere, Finland}

\bigskip

\centerline{\em $^2$Centre for Scientific Computing}
\centerline{\em P.O. Box 405, FIN - 02100 Espoo, Finland}

\bigskip
\centerline{\em $^3$Research Institute for Theoretical Physics}
\centerline{\em P.O. Box 9 (Siltavuorenpenger 20 C)}
\centerline{\em FIN - 00014 University of Helsinki, Finland}

\begin{abstract}
We present results of an extensive test program
of a group of pseudorandom number generators which are
commonly used in the applications of physics, in particular
in Monte Carlo simulations. The generators include public
domain programs, manufacturer installed routines and a random number
sequence produced from physical noise. We start by traditional
statistical tests, followed by detailed bit level and visual tests.
The computational speed of various algorithms is also scrutinized.
Our results allow direct comparisons between the properties of
different generators, as well as an assessment of the efficiency of
the various test methods. This information provides the best available
criterion to choose the best
possible generator for a given problem. However, in light of
recent problems reported with some of these generators, we also discuss the
importance of developing more refined physical tests to
find possible correlations not revealed by the present test methods.
\end{abstract}

\pagebreak
\pagenumbering{arabic}
\setcounter{page}{1}
%
%
\parskip=0.3cm
\renewcommand{\baselinestretch}{1.2}

%
%
%

\section{Introduction}

\begin{flushright}
{\small \em ``Have you generated any\\
              new random numbers today?''\\
{\sc J. K\"apyaho}}
\end{flushright}

Long sequences of random numbers are currently
required in numerous applications, in particular within statistical
mechanics,
particle physics, and applied mathematics.
The methods utilizing random numbers
include Monte Carlo simulation techniques  \cite{Bin92},
stochastic optimization \cite{Aar89}, and cryptography
\cite{Bri79,Dav89,Zhe91}, all of
which usually require
fast and reliable random number sources. In practice, the
random numbers needed for these methods
are produced by deterministic rules, implemented
as {\em (pseudo)random
number generator algorithms} which usually rely on simple aritmethic
operations. By their definition,
the maximum - length sequences produced by all these algorithms are
finite and reproducible, and can thus be ``random'' only in some limited
sense \cite{Com87,Com91}.

Despite the importance of creating good pseudorandom number
generators, fairly little
theoretical work exists on random number generation.
Thus, the properties of many generators are not well understood in depth.
Some random number generator algorithms
have been studied in the general context of cellular automata \cite{Com87},
and deterministic chaos \cite{Ber92}.
In particular, number theory has yielded exact results on the periodicity and
lattice structure for linear congruential
and Tausworthe generators \cite{Cov67,Mar68,Knu81,Tez87a}.
These results have led to theoretical methods of evaluating the
algorithms, the most notable being the so called
spectral test. However, most of these theoretical results are
derived for the full period of the generator while
in practice the behavior of subsequences of substantially shorter
lengths is of particular importance in applications. In addition,
the actual implementation of the random number generator algorithm
may affect the quality of its output. Thus, {\em in situ} tests
of implemented programs are usually needed.

Despite this obvious need for {\em in situ} testing of pseudorandom number
generators, only relatively few authors have presented
results to this end \cite{Lea73,Lec88,Pau84,Mar85}.
Most likely, there are two main reasons for this.
The first is the persistence of
underlying fundamental problems in the actual definitions of ``randomness''
and ``random'' sequences which have given no unique practical recipe
for testing a finite sequence of numbers \cite{Lam87}. Thus
various authors have developed an array of different tests which mostly
probe some of the statistical properties of the sequences, or test
correlations {\em e.g.} on the binary level. Recently, Compagner and Hoogland
\cite{Com87} have presented a somewhat more systematic approach to
randomness as embodied in finite sequences.
They propose testing the values of all possible correlation coefficients
of an ensemble of a
given sequence and all of its ``translations'' (iterated variations)
\cite{Com91b}, a task which
nevertheless appears rather formidable for practical purposes.
We are not aware of any attempts to actually carry out their program.
The second reason is probably more practical, namely
the gradual evolution of improved pseudorandom number generator
algorithms, which has led to a diversity of generators available in
computer software, public domain and so on. For many of these algorithms
(and their implementations) only a few rudimentary tests have been
performed.

In this work we have undertaken an extensive test
program \cite{Vat92} of a group of pseudorandom
number generators, which are often
employed in the applications of physics.
These generators which are
described in detail in Section 2 include
public domain programs GGL, RANMAR, RAN3, RCARRY and R250,
a library subroutine G05FAF,
manufacturer installed routines RAND and RANF, and even
a sequence generated from physical noise (PURAN II). Our strategy is to
perform a large set of different tests for all of these generators,
whose results can then be directly compared with each other.
There are two main reasons for this.
Namely, there is a difficulty associated with most quantitative
tests in the choice of the test parameters and final criteria for judging
the results. Thus, we think that
full {\em comparative} tests of a large group
of generators using {\em identical} test parameters and criteria can yield
more meaningful results, in particular
when there is a need for a reliable generator with a good overall performance.
Second, performing a large number of tests also allows a comparison
of exactly how efficient each test is in finding certain kinds of
correlations.

As discussed in Section 3, we first employ
an array of standard statistical tests, which
measure the degree of uniformity of the distribution of numbers,
as well as correlations between them. Following this, we perform
a series of bit level tests, some of which should be particularly
efficient in finding correlations between consecutive bits in
the random number sequences.
Third part of our testing utilizes visual pictures
of random numbers and their bits on a plane.
Finally, for the sake of completeness we
have also included a relative performance test of the generators in our
results. A complete summary of the test results is presented
in Section 4.
As our main result we find three generators, namely
GGL, G05FAF, and R250 with an overall best performance in all our
tests, although some other generators such as
RANF and RANMAR perform almost as convincingly.
We also find that the bit level tests are most efficient
in finding local correlations in the random numbers, but do
not nevertheless guarantee good statistical properties, as
shown in the case of RCARRY. Our results also show that
visual tests can indeed reveal spatial correlations not clearly detected in
the quantitative tests. Our work thus provides a rather comprehensive
test bench which can be utilized
in choosing a random number generator for
a given application.
However, choosing a ``good quality'' random number generator for {\em all}
applications may not be trivial as discussed in Section 5, in light
of the recent results reporting anomalous correlations
in Monte Carlo simulations \cite{Fer92,Gra93}
using the here almost impeccably performing R250.
Thus, more physical ways of testing
random number sequences are probably needed, a project which is
currently underway \cite{Vat93b}.

%
%
\pagebreak

%
%
%
\section{Generation of Random Numbers}

\begin{flushright}
{\small \em ``The generation of random numbers \\
              is too important to be left to chance.'' \\
{\sc R. Coveyou}}
\end{flushright}

Pseudorandom number sequences needed for high speed applications
are usually generated at run time using an algorithm
which often is a relatively simple nonlinear deterministic map.
The implementation of the
corresponding recurrence relation must also ensure
that the stream of numbers is reproducible from identical initial
conditions.
The deterministic nature of generation means that the designer
has to be careful in the choice of the precise relationship
of the recursion, otherwise unwanted correlations will appear as
amply demonstrated in the literature \cite{Com87}.

%
%
However, even the best generator algorithm can be defeated by a poor
computer {\em implementation}.
Whenever an exact mathematical algorithm is translated into
a computer subroutine, different possibilities for its implementation
may exist. Only
if the operation of a generator can be exactly specified on the
binary level, has the implementation a chance to be unambiguous;
otherwise, machine dependent features become incorporated into the routine.
These include finite precision of real numbers,
limited word size of the computer, and
numerical accuracy of mathematical functions.
Furthermore, it would often be desirable that
the implemented routine performed identically in each
environment in which it is to be executed, {\em i.e.} it would be
{\em portable}.

%
%
Some of the desired properties of good pseudorandom number generators
are easily defined but often difficult to achieve simultaneously.
Namely, besides good ``randomness'' properties
portability,
repeatability, performance speed, and a very long period are often required.
Ideally, a random number generator would be designed for each
application, and then tested {\em within that application} to ensure
that the inevitable correlations that do exist in a deterministic
algorithm, cause no observable effects.
In practice, this is seldom possible, which is another reason
why extensive tests of pseudorandom number generators are needed.

Most commonly used pseudorandom number generator algorithms are
the {\em linear congruential method},
the {\em lagged Fibonacci method},
the {\em shift register method},
and {\em combination methods}.
%
%
%
A special case are nonalgorithmic or {\em physical
generators} which are used for creating a non - reproducible sequence of
random numbers.
These are usually based on ``random'' physical events, {\em e.g.}
changes in physical characteristics of devices, cosmic ray bursts
or electromagnetic interference.
Details and properties of the algorithms will be summarized
in the next section. Following this, we shall describe in more
detail the particular generators chosen for our tests.
Reviews of current state of generation methods can be found in {\em e.g.}
Marsaglia \cite{Mar85}, James \cite{Jam90}, L'Ecuyer \cite{Lec90},
and Anderson \cite{And90}.

%
%
\subsection{Classification of Generation Methods}

\begin{flushright}
{\small \em ``Anyone who considers arithmethical methods \\
              of producing random digits is, of course, \\
              in a state of sin.'' \\
{\sc J. von Neumann}}
\end{flushright}

%
%

Among the simplest algorithms are the {\em linear
congruential generators} which
use the integer recursion
\begin{equation}
X_{i+1} = (a X_{i}\ +\ b\ )\mbox{ mod }m ,
\end{equation}
where the integers $a$, $b$ and $m$ are constants.
It generates a sequence $X_1, X_2, \ldots$ of random integers between
0 and $m-1$ (or in the case $b = 0$, between 1 and $m-1$).
Each $X_{i}$ is then scaled into the interval [0,1).
Parameter $m$ is often chosen to be equal or nearly equal to
the largest integer in the computer.
Linear congruential generators can be classified into {\em mixed}
$(b > 0)$ and {\em multiplicitive} $(b = 0)$ types, and
are usually denoted by LCG$(a,b,m)$ and MLCG$(a,m$), respectively.

Since the introduction of this algorithm by Lehmer \cite{Leh51}, its
properties have been researched in detail.
Marsaglia \cite{Mar68} pointed out about 20 years ago that the random
numbers in $d$ dimensions lie on a relatively small number of parallel
hyperplanes.
Further theoretical work \cite{Cov67,Fis82,Fis86}
 has been done to weed out bad choices
of the constants $a$, $b$ and $m$
but so far no consensus has evolved on a unique best choice
for these parameters.

%
%

To increase the period of the linear congruential algorithm,
it is natural to generalize it to the form
\begin{equation}
\label{lagg11}
X_{i} = (a_{1}\ X_{i-1}\ +\cdots +\ a_{p}\ X_{i-p})\mbox{ mod }\ m,
\end{equation}
where $p\ >\ 1$ and $a_{p} \neq 0$.
The period is the smallest positive integer $\lambda$ for which
\begin{equation}
\label{lagg22}
(X_{0},\ \ldots ,\ X_{p-1}) = (X_{\lambda},\ \ldots ,\ X_{\lambda+p-1}).
\end{equation}
Since there are $m^{p}$ possible $p\ -\ $tuples,
the maximum period is $m^{p} - 1$.
In this category the simplest algorithm is of the {\em Fibonacci type}.
The use of $p\ =\ 2,\ a_{1}\ = \ a_{2}\ = 1$ leads to the
Fibonacci generator
\begin{equation}
\label{lagg33}
X_{i} = (X_{i-1}\ +\ X_{i-2})\mbox{ mod }\ m.
\end{equation}
Since no multiplications are involved, this implementation has the
advantage of being fast.

A {\em lagged Fibonacci} generator requires an initial set of elements
$X_{1}, X_{2}, \ldots , X_{r}$ and then uses the integer recursion
\begin{equation}
\label{lagg44}
X_{i} = X_{i-r}\ \otimes\ X_{i-s},
\end{equation}
where $r$ and $s$ are two integer lags satisfying $r > s$ and $\otimes$
is a binary operation ($+$, $-$, $\times$, $\oplus$ (exclusive-or)).
The corresponding generators are designated by LF($r,s,\otimes$).
Typically, the initial elements are chosen as integers and the
binary operation is addition modulo $2^{n}$. Lagged Fibonacci
generators are elaborated in {\em e.g.} Ref. \cite{Mar85}.

%
%

An alternative generator type is the {\em shift register generator}.
Feedback shift register generators are also sometimes called
Tausworthe generators \cite{Tau65}.
The feedback shift register algorithm is based on the theory of primitive
trinomials of the form $x^{p} + x^{q} + 1$.
Given such a primitive trinomial and $p$ binary digits
$x_{0}, x_{1}, x_{2},\ldots , x_{p-1}$, a binary shift
register sequence can be generated by the following recurrence relation:
\begin{equation}
\label{r250e1}
x_{k} = x_{k-p}\ \oplus\ x_{k-p+q},
\end{equation}
where $\oplus$ is the exclusive-or operator, which is equivalent to
addition modulo 2.
$l$-bit vectors can be formed from bits taken from this binary sequence as
\begin{equation}
\label{r250e2}
W_{k} = x_{k}\ x_{k+d}\ x_{k+2d} \cdots x_{k+(l-1)d},
\end{equation}
where $d$ is a chosen delay between elements of this binary vector.
The resulting binary vectors are then treated as random numbers.
Such a generated sequence of random integers will have the maximum possible
period of $2^{p} - 1$, if $x^{p} + x^{q} + 1$ is a primitive trinomial
and if this trinomial
divides $x^{n} - 1$ for $n = 2^{p} - 1$, but for no smaller $n$.
These conditions
can easily be met by choosing $p$ to be a Mersenne prime, {\em i.e.}
a prime number $p$ for which $2^{p} - 1$ is also a prime.
A list of Mersenne primes can be found {\em e.g.} in Refs.
\cite{Zie68,Zie69,Bri79,Her92}.
%
%
Generators based on small values of $p$ do not perform well on the
tests \cite{Mar85}.
According to some statistical tests on computers \cite{Too71}
the value of $q$ should be small or close to $p/2$.

Lewis and Payne \cite{Lew73} formed $l$-bit words
by introducing a delay between the words.
The corresponding
generator is called the {\em generalized feedback shift register}
generator, denoted by GFSR$(p,q, \oplus)$.
In a GFSR generator the words $W_{k}$ satisfy the recurrence relation:
\begin{equation}
\label{r250e3}
W_{k} = W_{k-p} \oplus W_{k-p+q}.
\end{equation}
Under special conditions, maximal period length of $2^{p}-1$ can
be achieved.
Lewis and Payne \cite{Lew73} and Niederreiter \cite{Nie87}
have also studied the properties of the algorithm theoretically.
An important aspect of the GFSR algorithm concerns its initialization,
where $p$ initial seeds are required. This question has been studied
theoretically in Refs. \cite{Fus83a,Fus83b,Tez87a,Tez87b,Fus88}.

%
%
Given the inevitable dependencies that will exist in a pseudorandom sequence,
it seems natural that one should try to
{\em shuffle} a sequence \cite{Ito92} or to {\em combine}
separate sequences.
An example of such approach is given by MacLaren and Marsaglia \cite{Mac65}
who were apparently the first to suggest the idea of combining two
generators together to produce a single sequence of random numbers.
The essential idea is that
if $X_{1},\ X_{2},\ldots $ and $Y_{1},\ Y_{2},\ \ldots $ are
two random number sequences, then the sequence $Z_{1},\ Z_{2},\ \ldots $
defined by $Z_{i}\ =\ X_{i}\ \otimes\ Y_{i}$ will not only be more uniform than
either of the two sequences but will also be more
independent.
Algorithms using this idea are often called {\em mixed} or
{\em combination} generators.

%
%
%
As mentioned before, {\em physical devices}
have also been used in the creation of random number sequences.
Usually, however, such sequences are generated too slowly to be used
in real time, but rather stored in the computer memory where they
can be easily accessed. This also guarantees the reproducability
of the chosen sequence in applications. However,
physical memory restrictions often severely limit the
number of stored numbers. Unwanted and unknown physical correlations
may also affect the quality of physical random numbers.
As a result, physical random numbers have not been commonly
used in simulations.
One implementation of a physical generator can be found in Ref. \cite{Ric92}.

%
%
\subsection{Descriptions of Generators}

In this section, we shall describe in more detail the generators which have
been chosen for the tests. Since many combinations of possible parameters
exist, we have tried to choose those particular algorithms which have
been most commonly used in physics applications, or which have been
previously tested. At the end of this section, we shall also describe
a sequence of random numbers generated from physical noise, which has
been included for purposes of comparison.

%
%

\vspace*{0.5cm}
{\large \bf $\bullet$ GGL}

GGL is a uniform random number
generator based on the linear congruential method
\cite{Par88}.
The form of the generator is MLCG($16807,2^{31}-1$) or
\begin{equation}
\label{matlab}
X_{i+1} = (16807\ X_{i}) \mbox{ mod } (2^{31} - 1).
\end{equation}
This generator has been particularly popular
\cite{Par88}.
It has seen extensive use in the IBM computers \cite{IBM71}, and
is also available
in some commercial software packages such as subroutine RNUN in
the IMSL library \cite{IMSL89} and
subroutine RAND in the MATLAB software \cite{MATLAB91}.
MLCG($16807,2^{31}-1$) generators are quite fast and have
been argued to have good statistical properties \cite{Lew69}.
Results of tests with
and without shuffling are reported by Learmonth and Lewis \cite{Lea73}.
Other test results on implementations of this algorithm have been given
in \cite{Kir81,Alt88,And90,Lec90}.
Its drawback is its cycle length
$2^{31} - 1$ ($\approx 2 \times 10^{9}$ steps) \cite{Kir81},
which can be exhausted fast on a modern high speed computer.
We also note that our Fortran implementation of
GGL is particularly sensitive to the arithmetic accuracy
of its implementation (cf. Section 4). Our Fortran implementation of GGL
produces the same sequence as RNUN of the IMSL library
\footnote{We also unsuccessfully tried the IBM assembly code implementation of
Lewis {\it et al.} \cite{Lew69} on an IBM 3090 computer.}

%
%
\vspace*{0.5cm}
{\large \bf $\bullet$ RAND}

RAND uses a linear congruential random number algorithm
with a period of $2^{32}$ \cite{convex} to return successive pseudorandom
numbers in the range from 0 to $2^{31} - 1$.
The generator is LCG($69069,1,2^{32}$) or
\begin{equation}
\label{rand}
X_{i+1} = (69069\ X_{i}\ +\ 1) \mbox{ mod } 2^{32}.
\end{equation}
The multiplier 69069 has been used in many generators, probably
because it was strongly recommended in 1972 by Marsaglia \cite{Mar72}, and
is part of the SUPER - DUPER generator \cite{And90}.
Test results on various implementations of
the LCG($69069,1,2^{32}$) algorithm have been reported
in \cite{Lea73,Mar85,And90,Mar90b}.
The generator tested here is the implementation by Convex Corp. on
the Convex C3840 computer system \cite{convex}.

%
%
%
\vspace*{0.5cm}
{\large \bf $\bullet$ RANF}

The RANF algorithm uses two equations for generation of
uniform random numbers.
It utilizes the multiplicative congruential method with
modulus $2^{48}$.
The algorithms are MLCG($M_1,2^{48}$) and MLCG($M_{64},2^{48}$):
\begin{eqnarray}
X_{i+1} = (M_{1}\ X_{i}) \mbox{ mod } 2^{48}, \label{ranf1} \\
X_{i+64} = (M_{64}\ X_{i}) \mbox{ mod } 2^{48}, \label{ranf2}
\end{eqnarray}
where $M_1 = 44485709377909$ and $M_{64} = 247908122798849$.
Period length of the RANF generator is $2^{46}$ \cite{Mat90}.
Spectral test results on the RANF generator
have been given in Refs. \cite{And90,Fis90}.
On the CRAY-X/MP and CRAY-Y/MP systems,
RANF is a standard vectorized library
function \cite{Cray}.
The operations $(M_{1} X_{i})$ and $(M_{64} X_{i})$
are done as integer multiplications
in such a way as to preserve the lower 48 bits.
We tested RANF on a Cray X-MP/432.

%
%
%
\vspace*{0.5cm}
{\large \bf $\bullet$ G05FAF}

G05FAF is a library routine in the NAG software package \cite{NAG90}.
It calls G05CAF which is a multiplicative congruential algorithm
MLCG($13^{13},2^{59}$) or
\begin{equation}
\label{g05faf}
X_{i+1} = (13^{13}\ X_{i})\mbox{ mod } 2^{59}.
\end{equation}
G05FAF can be used to generate a vector of $n$ pseudorandom numbers
which are exactly the same as $n$ successive calls to the G05CAF routine.
Generated pseudorandom numbers are uniformly distributed
over the specified interval $[a,b)$.
The period of the basic generator is $2^{57}$ \cite{NAG90}.
Its performance has been analyzed by the spectral test \cite{Knu81}.

%
%
%
\vspace*{0.5cm}
{\large \bf $\bullet$ R250}

R250 is an implementation of a generalized feedback shift
register generator \cite{Lew73}.
The 31-bit integers are generated by a recurrence of the form
GFSR($250,103,\oplus$) or
\begin{equation}
\label{shiftreg}
X_{i} = X_{i-250}\ \oplus\ X_{i-(250-103)}.
\end{equation}

Implementation of the algorithm is straightforward, and
$p=250$ words of memory are needed to store the $250$ latest random numbers.
A new term of the sequence can be generated by a simple exclusive - or
operation.
An IBM assembly language
implementation of this generator has been presented by
Kirkpatrick and Stoll \cite{Kir81} who use a MLCG($16807,2^{31}-1$)
to produce the first $250$ initializing integers.
Due to the popularity of R250, there have been many different
approaches for its initialization \cite{Pay78,Fus83a,Col86,Fus88}.
The period of the generator is $2^{250} - 1$ \cite{Kir81}.
Some test results of R250 generator have been reported by Kirkpatrick and
Stoll \cite{Kir81}.
We have implemented R250 on Fortran \cite{Hel85}.

%
%
%
\vspace*{0.5cm}
{\large \bf $\bullet$ RAN3}

RAN3 generator is a lagged Fibonacci generator LF(55,24,$-$) or

\begin{equation}
		X_i = X_{i-55} - X_{i-24}.
\end{equation}

The algorithm has also been called a subtractive method.
The period length of RAN3 is $2^{55} - 1$ \cite{Knu81}, and
it requires an initializing sequence of $55$ numbers.
The generator was originally Knuth's suggestion
\cite{Knu81} for a portable routine but with an add operation
instead of a subtraction. This was translated to a real Fortran
implementation by Press {\it et al.} \cite{Pre89}.
We were unable to find any published test results for RAN3.

%
%
%
\vspace*{0.5cm}
{\large \bf $\bullet$ RANMAR}

RANMAR is a combination of two different generators \cite{Jam90,Mar90a}.
The first is a lagged Fibonacci generator
\begin{equation}
\label{ranm11}
X_{i}  = \left\{\begin{array}{ll}
          X_{i-97} - X_{i-33}, & \mbox{if $X_{i-97} \geq X_{i-33};$} \\
           X_{i-97} - X_{i-33} + 1, & \mbox{otherwise.} \\
      \end{array}
     \right .
\end{equation}
Only 24 most significant bits are used for single precision reals.
The second part of the generator is a simple arithmetic sequence for the
prime modulus $2^{24} - 3 = 16777213$.
The sequence is defined as
\begin{equation}
\label{ranm22}
  Y_i = \left\{\begin{array}{ll}
        Y_i - c,   & \mbox{if $Y_i\geq c$;} \\
        Y_i - c + d, & \mbox{otherwise,} \\
                            \end{array}
                  \right .
\end{equation}

where $c = 7654321/16777216$ and $d= 16777213/16777216$.

The final random number $Z_i$ is then produced by combining the obtained
$X_i$ and $Y_i$ as

\begin{equation}
Z_{i}  = \left\{\begin{array}{ll}
          X_{i} - Y_{i}, & \mbox{if $X_{i} \geq Y_{i};$} \\
           X_{i} - Y_{i} + 1, & \mbox{otherwise.} \\
      \end{array}
     \right .
\end{equation}

The total period of RANMAR is about $2^{144}$ \cite{Mar90a}.
A scalar version of the algorithm
has been tested on bit level with good results \cite{Mar90a}.
 We used the implementation by James \cite{Jam90}
which is available in the Computer Physics Communications (CPC)
software library, and has been recommended for a
universal generator.

%
%
%
\vspace*{0.5cm}
{\large \bf $\bullet$ RCARRY}

RCARRY \cite{Mar90b} is based on the operation known
as ``subtract - and - borrow''.
The algorithm is similar to that of lagged Fibonacci, but it has the
occasional addition of an extra bit.
The extra bit is added if the Fibonacci sum is greater than one.
The basic formula is:
\begin{equation}
\label{rcarry}
X_{i} = (X_{i-24}\ \pm\ X_{i-10}\ \pm c) \mbox{ mod } b.
\end{equation}
The carry bit $c$ is zero if the sum is less than or equal to $b$,
and otherwise ``$c=1$ in the least significant bit position''
\cite{Jam90}. The choice for $b$ is $2^{24}$.

The period of the generator is about $2^{1407}$ \cite{Mar90b}
when 24-bit integers are used for the random numbers.
We were unable to find any published test results for RCARRY.
We used the implementation of James \cite{Jam90},
again available in the CPC software library.

%
%
%
\vspace*{0.5cm}
{\large \bf $\bullet$ PURAN II}

PURAN II is a physical random number generator created by
Richter \cite{Ric92}.
It uses random noise from a semiconductor device.
The generated data has been permanently
stored on a computer disk, from which it can
be transferred by request.
In this work, we have tested the PURAN II data on bit level only
(cf. Section 4), and also used it to verify the
correct operation of our test programs.

%
\pagebreak

\section{Description of the Tests}

\begin{flushright}
{\small \em ``Of course, the quality of a generator can \\
              never be proven by any statistical test.'' \\
{\sc P. L'Ecuyer}}
\end{flushright}

A fundamental problem in testing finite pseudorandom
number sequences stems from the fact that the definition of  randomness
for such sequences is not unique \cite{Lam87}. Thus, one usually
has to decide upon some criteria which test at least the most
fundamental properties
that such sequences should possess, such as correct values of the
moments of their probability distribution. This has lead to the emergence
of a large number of tests which
can be divided into three approximate categories:
{\em Statistical (or traditional) tests} for testing random numbers in
real or integer representation,
{\em bit level tests} for binary representations of random numbers,
and more phenomenological {\em visual tests}. In this work, we have employed
several tests belonging to each of these categories, as will be discussed
below. Also, the spectral test for LCG generators was included.
We should note here that recently Compagner and Hoogland \cite{Com87}
have suggested a more systematic test program for finite sequences.
We shall not employ it in this work, however.

The traditional utilitarian approach has been
to subject pseudorandom number sequences to
tests, which derive from mathematical statistics \cite{Knu81}.
In their simplest form, tests in this category reveal possible
deviations of the distribution of numbers from an uniform distribution,
such as the $\chi^2$ test. However, some of the
more sophisticated tests should actually probe correlations between successive
numbers as well \cite{Lec88}.

Another approach is to test the properties of random numbers
on the bit level. Of the traditional tests, some can be performed
in this manner also. Marsaglia \cite{Mar85} has proposed additional
tests which explicitly
probe the individual bits of random number sequences
represented as binary computer words.
Some of these tests have been further refined \cite{Alt88}. We have
included two of these tests here, in particular to examine
possible correlations
between bits of successive binary words.

A rather different way of testing spatial correlations between
random numbers is possible by using direct visualization. This can
most easily be done in two dimensions by plotting pairs of points
on a plane, or visualizing the bits of binary numbers. In addition to
yielding qualitative information, such tests offer a possibility to
develop more physical quantitative tests through interpretation
of the visualized configurations as representations of physical
systems, such as the Ising model \cite{Com87,Vat93b}. In this work,
however, we have simply used a few different types of visual
tests to complement our quantitative tests.

Before discussing each test in detail, we would like to emphasize that
although some of the generators we have tested have previously been
subjected to similar tests,
an extensive comparative testing of a large {\em collection}
of generators has been lacking up to date.
The importance of this becomes obvious when one
considers the freedom of choice of various parameters in the tests, as
discussed below.
Only comparative testing with identical parameters
allows a direct comparison between different generators.
Another difficulty concerns the implementation of random number generator
algorithms and the testing routines
\cite{Sch79,Par88,And90,Gen90}. Problems in either may actually
lead to significant differences in the results.  In fact, as an
example we shall explicitly demonstrate for GGL and RAND
how slightly different
implementations of the same generator can lead to completely different
results.


\subsection { Statistical tests}

The statistical
tests included in our test bench were the uniformity test, the serial test,
the gap
test, the maximum of $t$ test, the collision test, and the run test.
In addition, we carried out the park test \cite{Mar85}.
A review of the statistical tests can
be found {\em e.g.} in Ref. \cite{Knu81}, and a suggestion for implementing
them in Ref. \cite{Dud81}.

The leading idea in carrying out the statistical tests was to
improve the statistical accuracy of these tests by utilizing
a one way Kolmogorov - Smirnov (KS) test. This was achieved by repeating
each individual test described below
$N$ times, and then submitting the obtained
empirical distribution to a KS test (for the park test, however,
this was not possible).
Similar approach has been suggested earlier by Dudewicz
and Ralley \cite{Dud81} and realized by L'Ecuyer \cite{Lec88}.
The KS test reveals deviations of an empirical
distribution function ($F_n(x)$) from the theoretical one ($F(x)$).
This can be quantified by test variables $K^+$ and $K^-$, which
are defined by $K^{+} = \sqrt{n} \sup \{F_{n}(x)-F(x)\} $ and
$K^{-} = \sqrt{n} \sup \{F(x)-F_{n}(x)\} $.
$K^+$ measures the maximum deviation of $F_n(x)$ from
$F(x)$ when $F_n(x) > F(x)$ and $K^-$ measures the respective
quantity for $F_n(x) < F(x)$. The tests are as follows:

\newcounter{stat1}
\begin{list}%
{(\roman{stat1})}{\usecounter{stat1}}

\item  To test the uniformity of a random number sequence,
       a standard $\chi^2$ test was used \cite{Knu81}. $n$ random numbers
	were generated in the  half open interval $[0,1)$, then
        multiplied by $\nu$ and truncated to integers
	in the interval $[0,\nu )$.
	The number of occurrences in each of the $\nu$ bins was compared
	to the theoretical prediction using the $\chi^2$ test.

\item Serial correlations were tested \cite{Lea73,Knu81}
	by studying the occurrence of
       $d$-tuples of $n$ random numbers distributed in the interval $[0, 1)$.
	For example, in the case of pairs, we tabulated the
	number of occurrences of $(x_{2i},x_{2i+1})$ for all
	$i$ $\in$ $[0,n)$.
	Each $d$-tuple occurs
	 with the probability $\nu ^{-d}$ where $\nu $ is the number of bins
	in the interval. The results were then
	subjected to the $\chi^2$ test.

\item The gap test \cite{Knu81} probes the uniformity of the random number
	    sequence of length $n$. Once a random
            number $x_i$ falls within a given interval $[\alpha,\beta]$,
	    we observe the number of subsequent numbers
            $x_{i+1}\ldots x_{i+j-1} \not\in [\alpha,\beta]$. When
	    again $x_j\in [\alpha,\beta]$, it defines
	    a gap of length $j$. For finite sequences, it is
            useful to define a maximum gap length $l$.
            Then we can test the results against the
	    theoretical probability using
	    the $\chi^2$ test.

\item      The maximum of $t$ test \cite{Knu81} is a simple uniformity test.
	   If we take a random number sequence of length $n$
	   ($x_i \in [0,1), i= 1, \ldots ,n$)
           and divide it into subsequences of length $t$
	   and pick the  maximum value for each subsequence,
	   the  maxima should follow the $x^t$ distribution.

\item      The collision test \cite{Knu81} can be used to test the
	   uniformity of the sequence when the number of available
	   random numbers ($n$) is much less than the number of bins ($w$).
	   We then study how many times a random number falls in the same bin,
	   {\em i.e.} how many collisions occur. The probability for
           $j$ collisions is:

\begin{equation}
 {{w(w-1) \cdots ( w- n + j + 1)} \over {w^{n}}} {n \choose {n - j}},
\end{equation}

where $w = s^d$, $d$ is the dimension and $s$ can be chosen.

\item 	In the run test \cite{Lea73,Knu81}, we calculate the number of
	occurrences of increasing or decreasing subsequences
        of length $1 \le i < l$  for a given
	random number sequence ${x_1,x_2,\ldots ,x_n}$.
        To carry out this test we chose $l=6$ and
 	followed Knuth \cite{Knu81} in the choice of the
        relevant test quantity.

\item   In the park test \cite{Mar85}, we choose randomly points in a
        $d$-dimensional space and
	allocate a diameter for each point. Within each diameter,
	``a car is parked''. The aim is to park as many
	non - overlapping cars as possible, and study the
	distribution of	$k$ cars. Unfortunately, since the
        theoretical distribution is not known, this test can only be
        used for qualitative comparative studies.

\end{list}


\subsection {Bit level tests}

Two of the tests included in the previous section on statistical tests,
namely the run test and the collision test could
equally well be included in the category of bit level tests as
they can also be performed for binary representations of random numbers.
Recently, Marsaglia \cite{Mar85} has introduced new tests in his
DIEHARD random number generator test bench.
Of these we carried out the $d$-tuple and the rank tests.
We shall briefly describe both of them below.

\newcounter{bit1}
\begin{list}%
{(\roman{bit1})}{\usecounter{bit1}}

\item 	The $d$-tuple test realized here is a modified version \cite{Alt88}
	of the original \cite{Mar85}. We extended the test by improving its
	statistical accuracy by submitting the empirically obtained
	distribution to a Kolmogorov - Smirnov test.
	In the $d$-tuple test, we represent a random integer $I_i $
	as a binary sequence of $s$ bits $b_{i,j}, (j = 1,\ldots ,s)$:
\begin{eqnarray}
I_{1} & = & b_{1,1} b_{1,2} b_{1,3} \cdots b_{1,s}, \nonumber \\
I_{2} & = & b_{2,1} b_{2,2} b_{2,3} \cdots b_{2,s}, \nonumber \\
&   & \vdots \nonumber \\
I_{n} & = & b_{n,1} b_{n,2} b_{n,3} \cdots b_{n,s},
\end{eqnarray}
	where an obvious choice for the parameter $s=31$
	(in testing RAN3, we used $s=30$).
	Each of the binary sequences $I_i$ is divided into
	subsequences of length $l$ which can be used to form
	$n$ new binary sequences $I'_i=b_{i,1}b_{i,2}\ldots b_{i,l}$.
	These sequences are then joined into one more binary
	sequence of length $d\times l$
	in such a way that these final sequences
	$\bar I_i$ partially overlap:
\begin{eqnarray}
\bar{I}_{1} & = & b_{1,1} \cdots b_{1,l} b_{2,1} \cdots b_{2,l}
\cdots b_{d,1} \cdots b_{d,l}, \nonumber \\
\bar{I}_{2} & = & b_{2,1} \cdots b_{2,l} b_{3,1} \cdots b_{3,l}
\cdots b_{d+1,1} \cdots b_{d+1,l}, \nonumber \\
&   &   \vdots \nonumber \\
\bar{I}_{n} & = & b_{n,1} \cdots b_{n,l} b_{n+1,1} \cdots b_{n+1,l}
\cdots b_{n+d-1,1} \cdots b_{n+d-1,l} .
\end{eqnarray}
	Each of these new integers falls within $\bar I_i \in [0,2^{dl}-1]. $
	In the test, the values $\bar I_i$ of the new random numbers are
	calculated as well as the number of respective occurrences.
	A statistic which follows the $\chi^2$ distribution can be
	calculated although the subsequent sequences are correlated
	\cite{Alt88}. The $N$ results of the $\chi^2$ test
	were finally subjected to a KS test.

\item	For the rank test, we construct a $(v \times w)$
        random binary matrix from the random
	numbers. The probability that the
	rank $r$ of such a matrix equals
	$r = 0,1,2,\ldots , \min(w,v)$ can be calculated \cite{Mar85}
	allowing us to perform the $\chi^2$ test, followed by
	a KS test.

\end{list}


\subsection {Spectral test}

The spectral test was included for the sake of completeness. Unlike
other tests, it relies on the theoretical properties of LCG algorithms
{\em independent} of their implementation. It
has been used extensively to characterize
the properties of linear congruential generators
\cite{Cov67,Knu81,Fis82,Fis86,Lec88,Par88,And90,Fis90}, mainly
in order to find ``good'' values for the parameters within them.
It probes the maximal distance between hyperplanes on which the
random numbers produced by an LCG generator fall \cite{Mar68}.
The smaller the distance, the ``better'' the generator.

All the linear congruential generators used in this study were
subjected to the spectral test.
We used two figures of merit \cite{Knu81}, namely
\begin{equation}
\kappa_d = {\nu_{d} \over {\gamma_d m^{1 \over d}}},
\end{equation}
where $d$ is dimensionality and
$m$ is the period of the LCG generator
in question. The wave number $\nu_d$ is the inverse of the
maximal distance between the hyperplanes in $d$ dimensions, and
the coefficient $\gamma_d$ depends on dimension and is tabulated
{\em e.g.} in Knuth \cite{Knu81}. Basically, the denominator is the inverse of
the theoretical minimal distance between hyperplanes \cite{Lec88}, and
thus $\kappa_d$ is the normalized distance between hyperplanes.
The other figure of merit is
\begin{equation}
\lambda_d = \log_2(\nu_d),
\end{equation}
which gives the number of bits uniformly distributed in $d$ dimensions
\cite{Knu81,And90}. The larger $\lambda_d$, the ``better'' the generator.


\subsection{Visual tests}

Visual tests can provide additional qualitative information about
the properties of random
number generators and can further corroborate the results of quantitative
tests. We submitted the generators to four visual tests:

\newcounter{vis1}
\begin{list}%
{(\roman{vis1})}{\usecounter{vis1}}

\item 	The distribution of random number pairs was plotted in two dimensions
	to see if there exists any ordered structures. For an LCG generator,
	one should be able to distinguish the hyperplanes on which the random
	numbers fall \cite{Mar68,Knu81}. The shorter the interplanar distance,
	the ``better'' the generator.
	Also lagged Fibonacci generators as well as
	shift register ones are known to produce some structure which should be
	visible with some choices of parameters \cite{Com87,And90}.

\item 	To study binary sequences visually one can plot
	the random numbers as binary computer words on a plane.
	Ones were mapped onto black squares and zeros onto white ones.
        The consequent figure can also be interpreted as a configuration
        of a two dimensional Ising model at an infinite
        temperature which could be subjected to quantitative tests
        \cite{Vat93b}.

\item 	First $n$ random numbers were generated. Then the distance
	$\vert x_i - x_j \vert$ from $x_i$ was calculated for
	all $j = 1,\ldots ,n $. This was done for all $x_i$
	 ($i = 1,\ldots ,n $), and then the distance was plotted
	in two dimensions with gray scale colors: the
	lighter the color, the larger the difference. Areas of uniform gray
	shade indicate possible
	local correlations in the random number sequence.

\item 	The gap test was visualized by calculating the difference
	$\vert x_i - x_j \vert$, where $i \in [\alpha,\beta]$ and
	$j=1,\ldots ,l$ where $l$ is the maximum gap length as in section 3.1.
	The difference was plotted in gray shade colors as in (iii) above.
	Uniform darkening or lightening of gray shades indicates a gap.

\end{list}

\newpage

\section{Results}

\begin{flushright}
{\small \em ``A random number generator is much like sex: \\
              when it's good it's wonderful, \\
              and when it's bad it's still pretty good.'' \\
{\sc G. Marsaglia}}
\end{flushright}

The random number generators were initialized
with the seed 667790
\linebreak
($ = 10100011000010001110_2$)
except for R250 which was initialized with GGL
(to 24 significant bits) using this seed. Following initialization,
consequtively generated sequences of
random numbers were subjected to statistical
and bit level tests which were repeated $N$ times, after which the one way
Kolmogorov - Smirnov test was applied
to the $N$ results to further improve the statistics.
Thus, the final test variables are the $K^-$ and $K^+$ values.
The only exception is the maximum of $t$ test where an additional
KS test was applied to the results of the first KS test.
A generator was considered to fail a test if the descriptive level
$\delta^-$ or $\delta^+$ was less
than 0.05 \cite{Lec88} or larger than 0.95.
In other words, a failure occurred if the
empirical distribution followed too closely or was too far from the
theoretical one.
We note that to verify the independence of the results on the choice
of the seed, two other choices for initial seeds for RAND and RAN3 were tested
namely 1415926535 ($=1010100011001010101001100000111_2$) (from the decimals
of $\pi$), and $2^{15}$ ($= 32768 = 1000000000000000_2$).
No changes in the results of the $d$-tuple test were found.
Our tests were performed on a Convex C3840, a Silicon Graphics
Iris 4D380 VGX and a Kubota Titan 3000. A Cray X-MP was used for testing RANF.

\subsection{Standard tests}

The parameters used in the standard tests are given in Table
\ref{tab:stand_param}. The numbering
refers to  Table \ref{tab:standard}
where the results are shown. In the choice of parameters, we
followed  L'Ecuyer \cite{Lec88}, with some changes
to improve the statistical accuracy of the results.
If a generator failed a given test,
it was subjected to another test starting from the state it
reached in the first test.
If a second failure occurred, one more test was
performed by starting from a new initial state with the seed
14159 ($ = 11011101001111_2$).

In Table \ref{tab:standard}, frames with thin lines
indicate a single failure, frames with double single lines two
failures, and frames with bold lines three consequtive failures in the
corresponding tests.
Additionally, as overall ``goodness'' factors
for each generator, we have calculated
relative deviations $R_{\delta}$ and $R_K$
from the average theoretical descriptive level values $\delta^+$ and
$\delta^-$ (shown in the table) and the corresponding KS test values
$K^+$ and $K^-$, respectively. They are
shown at the bottom of the table.
Based on our results, the performance of the generators falls in three
rough categories. GGL, R250, RANF, and G05FAF all display only single
failures, RAND and RAN3 fail two consecutive times, and RANMAR and
RCARRY fail there consequtive times at least once.
The calculated goodness values support these results, too.
The performance of RCARRY is noticeably poor in the
gap test which suggests possible local correlations
in the random number sequence.
We should note that although the calculated $R$ values give an indication of
the overall performance,
it is clear that one should
be aware of the particular weaknesses of each generator before a
specific application is considered.
Finally, as a qualitative counterpart to the statistical tests,
the park test was carried out with two different ``car sizes'',
{\em i.e.} $d=0.001$, $n = 10^6$, and $d=0.01$, $n=10^5$.
However, we found the results for all generators to agree within errors.
Thus, the test gave no additional information in the present case.

When comparing our results with the literature,
one should note that the
actual implementations of the generators tested may differ from ours.
Different implementations of the same algorithm may change the generated
random number sequence, with unknown consequences for its properties.
This makes direct comparisons of the results difficult. Another
problem lies in the possible machine dependence of a ``bad'' implementation.
An example of this is GGL, which when implemented in single precision mode
in 32 bit computers gives a period of 32, as will be discussed
later in the context of the visual test results.

Various implementations of the algorithm MLCG($16807,2^{31}-1$)
(GGL here)
have previously been tested extensively, see {\em e.g.} Refs.
\cite{Lew69,Lea73,Gar78,Kir81}. In particular,
Lewis {\em et al.}  \cite{Lew69}
performed a series of tests
when they introduced the IBM SYSTEM/360
assembly language implementation of MLCG($16807,2^{31}-1$).
They used sequences of lengths
$n=2^{16}+5 = 65541$ with tests repeated ten times.
The authors characterized the quality of the results by the maximum standard
deviations
$\sigma_{\rm max}$ from the mean.
In the uniformity test, the sequence was divided into $\nu =2^{12} = 4096$
bins, and all results were within $\sigma_{\rm max}=1.9$ from the mean.
The serial test for pairs $(x_i,x_{i+l})$
was performed with various lags $l$
and number of bins
$\nu = 256$, with results within $\sigma_{\rm max}=2.3$.
Furthermore, the run test was completed with $l=8, n = 65541$,
and repeated ten times, with ``much larger'' standard deviations.
However, Lewis {\em et al.}
concluded that the run test together with their other tests
gave ``no evidence of departures from randomness'' \cite{Lew69}.
In another reference \cite{Lea73},
an unspecified implementation of the generator
has also been subjected to the run test with a two way
KS test and has been found to pass it, as well as a serial test
for triples ($d=3$) but to fail a serial test for
pairs ($d=2$). In the run test, the sequence length was
$n=65536$, the runs were counted up to $l=8$ and the test
was repeated one hundred times ($N=100$). Similarly, in the serial test
$n=65541$ and $N=100$ \cite{Lea73}.
We should note that in our tests, where better
statistics was used ($N=1000$) GGL passed the run test
and all the serial tests for $d=2,3,4$.

RAND is an implementation of LCG($69069,1,2^{32}$) by Convex Corp.
\cite{convex}.
Learmonth and Lewis \cite{Lea73} have also tested their assembly
language implementation
of LCG($69069,1,2^{32}$) with the same test as the
MLCG($16807,2^{31} -1$) generator discussed above.
LCG($69069,1,2^{32}$) passed the serial test
for both $d=2$ and $d=3$, as well as the run test.
In our tests, RAND passed all these tests as well.

An IBM assembly language implementation of
 GFSR($250, 103, \oplus $) has been found to perform well
in a run test with parameters
$n = 10^6, N=30, l=9$ \cite{Kir81}.
Our Fortran implementation of it, namely R250 also passed the run test.

Finally, we have been unable to find any published data on
statistical tests for any implementations of RANMAR,
RCARRY, RAN3, RANF or G05FAF.


\subsection{Spectral test}

The results of the spectral test for the LCG generators
of this paper are presented in Table \ref{tab:spectral}.
In the case of RANF, we show results for the two generators
which comprise it, namely MLCG(44485709377909,$2^{48}$) (RANF1) and
\linebreak
MLCG(247908122798849,$2^{48}$) (RANF2). Overall,
the generator G05FAF from the NAG library is the most successful in the
test. On the other hand,
GGL displays the known flaw of this generator performing poorly
at low dimensions. All generators, however, gave worse results
in most dimensions than the
minimum acceptable values suggested by Fishman and Moore
\cite{Fis86,And90}. We note that since the results of this test
are independent of the implementations of the algorithms, our
results for RAND agree with previous results for
LCG($69069,1,2^{32} $) \cite{And90}, for GGL with
MLCG($16807,2^{31}-1$) \cite{Fis86,Lec88,And90} and for RANF1 with
MLCG(44485709377909,$2^{48}$) \cite{And90,Fis90}.


\subsection{Bit level tests}

The bit level tests probe the properties of the individual
bits which comprise the random numbers
thus testing properties somewhat different from the statistical
tests. We chose to
use the generalized $d$-tuple test and the rank of a random matrix test
for studying correlations on the bit level.

The $d$-tuple test ($d=3$) was carried out for $n = 5000$ random numbers,
each of which was coded into a $31$ bit binary sequence
(for RAN3, the sequence length was 30). Of this
sequence, we chose bit strips of width $l=3$. The $\chi^2$
test was performed $N = 1000$ times, and the results were
then subjected to a KS test\footnote{Results of a systematic study
indicate \cite{Vat93b} that with these
parameters, the $d$-tuple test can detect correlations
up to about fifty numbers apart.}.
This test was performed twice
for each generator (excluding PURAN II) and
we considered ``failed'' only those bits that failed twice in succession.

The rank test was carried out with parameters
$v = w = 2, n= 1000$ and $N=1000$. The $(2\times 2)$
random matrices were formed
systematically using the $i^{\rm th}$ and $(i+1)^{\rm th}$
($1 \le i \le 31$) bit pairs from each two successive numbers.
The test was performed twice to all
pseudorandom number generators with the same failing criteria as in
the $d$-tuple test.

Results for the
$d$-tuple and rank tests are shown in Table \ref{tab:bitcorr1}.
Details of the implementations and initializations of the
generators are also shown there. In our notation, the
bit number one is the most significant bit.
GGL, G05FAF, and R250 pass both tests with an impeccable
performance, in that none of the 31 bits show observable
correlations. The physically generated random numbers of PURAN II
also pass both tests.
For RANMAR and RCARRY, only the
24 most significant bits are guaranteed to be good \cite{Jam90}, which
our tests confirm. On the other hand,
RAND and RAN3 show significant correlations. In particular,
the correlations
in RAN3 are serious since they affect the five most significant bits.
When RAND was called in integer form, it gave one more correlated
bit than the calls in floating point representation both in the
$d$-tuple and rank tests.

Previously, an unspecified
implementation of MLCG($16807,2^{31}-1$)  (GGL here) has been shown
to pass a simpler version of the $d$-tuple test ($d=l=3, n=2000$, test
repeated five times) \cite{Alt88}. An unspecified implementation of
LCG($69069, 1, 2^{32}$) (RAND here) failed the same test with 11 failing bits
\cite{Alt88}.
An IBM assembly language implementation of
GFSR($250, 103, \oplus $) (R250 here)
and an unspecified implementation of
MLCG($16807,2^{31}-1$) (GGL here) have been shown to
pass a test which probed possible correlations in the
five most significant bits by studying triples of random
numbers by placing them on a unit cube with a resolution of
$32 \times 32 \times 32$ cells
($n=10^6$ and the test was repeated ``several'' times) \cite{Kir81}.
Kirkpatrick and Stoll \cite{Kir81}
further argue that as all columns of bits generated
by a GFSR($250, 103, \oplus $) generator have the same statistical
characteristics, their results of this test should apply to any subset of
bits in a random number sequence produced by this generator. This is
in accordance with our results, where no correlations were found in the 31
bits of R250 (see Table \ref{tab:bitcorr1}).

The initialization of
R250 deserves a more detailed discussion. Already Kirkpatrick
and Stoll \cite{Kir81} have pointed out
that the algorithm
GFSR(250,103,$\oplus $) requires a careful initialization.
Our results in Table \ref{tab:bitcorr2}
clearly show this to be true as the results for R250
initialized with RAN3 show correlations in the most significant
bits, an obvious consequence of the bad quality of RAN3.
It is particularly important to notice, that these correlations
once present seem to persist in R250.

Finally, for testing purposes we also realized our own Fortran
implementation of LCG($69069,1,2^{32}$)
in double precision accuracy. In this implementation,
whenever the sign bit equalled one it was flipped to zero. Thus, the
random numbers remained between  zero and $2^{31}-1$. This
implementation produced exactly the same sequence as
RAND on a Convex C3840.
Another possible implementation of the same algorithm
was then realized in such a way that whenever the sign bit equalled one,
the whole computer word was shifted to the right (with periodic boundary
conditions) until a zero was obtained for the sign bit.
When bit level tests were done for this implementation, all 31 bits
failed. This dramatically highlights the effect of a poor implementation
on the performance of the same algorithm.


\subsection{Visual tests}

The two dimensional distribution of 20 000 random number pairs
$(x_i,x_{i+1})$ from GGL, RAND and R250 is shown in Figs. 1$(a)-(f)$.
When plotted on the scale from zero to one (Figs. 1$(a),(c),(e)$),
no generator shows any discernible structure.
However, when the 20 000 random numbers
are plotted on an expanded scale (Figs. $1(b),(d),(f)$)
one can clearly see the random numbers ordering
on planes in the cases of GGL and RAND.
This kind of behavior is expected for LCG generators \cite{Mar68},
and the results are in accordance with the spectral test of Section
3.3. We note that no structure on other generators was observed on
this scale.

In Figs. 2$(a)$ and $(b)$, we depict subsequent random numbers in
binary form on a $124 \times 124 $ matrix from our best implementations of
GGL and R250, respectively. Although the former showed clear lattice
structure in the test above, in binary form it is very difficult to
find any differences between these two generators. More quantitative
tests of the bit maps shown here are in progress \cite{Vat93b}.
When we  further compare the binary representations of R250 with different
initializations, the visual tests corroborate
the findings of the qualitative tests: In Fig. 2$(b)$, R250 is
initialized with GGL with double precision modulo operation, returning
integers.
However, if we initialize R250 with real numbers from
GGL implemented in single precision, the result
is catastrophic as seen in Fig. 3. Clearly this is an improper way
to implement MLCG($16807,2^{31}-1$) in 32 bit word computers.
It is interesting to note,
that already Lewis {\em et al.} \cite{Lew69} pointed out the need to
use double precision accuracy
with the assembly implementation of MLCG($16807,2^{31} -1$).

The visualization of the difference between random numbers (test (iii))
and the gap test visualization (test (iv))
gave rather inconclusive results and thus yielded
no further insight to the properties of the generators.

Finally, a problem was encountered with the decoding
program which was included with the physical PURAN II
random numbers \cite{Ric92}.
When used to extract random numbers in floating
point representation, we found that it produced numbers which
fell on planes similarly to the linear congruential
generators, although PURAN II passed all bit level tests.
However, when using the decoding algorithm in integer
format the problem disappeared.


\subsection{Speed of generators}

We tested the computational speed of the eight generators both on
a Cray X-MP/432 and a Convex C3840. All generators were
compiled in two ways:
first, only scalar optimization was allowed and
second, also vectorization was allowed. The testing was done for
sequences of  lengths $n=1, 10, 100, 1000, 10 000$ and $100 000$.

Results are in Table \ref{tab:speed} for $n=1$ and
$n=1000$ in units of microseconds ($\mu$s) per random number call.
The speedup for longer sequences ($n > 1000$)
per random number call is nonexistent.
First,
Cray's own generator RANF was
always the fastest on it which indicates a successful
implementation in this sense.
Other generators are almost equally fast
for short sequences, except for R250. On the other hand,
the performance for longer sequences is fastest for R250 and G05FAF if
vectorizing is allowed.
The code for RAN3 as given in Numerical Recipes \cite{Pre89}
was incompatible with Cray and is thus omitted from its
performance results.

\newpage

\section{Summary and Discussion}

\begin{flushright}
{\small \em ``The whole history of \\
              pseudorandom number generation \\
              is riddled with myths and extrapolations\\
              from inadequate examples. \\
              A healthy dose of sceptisism is needed \\
              in reading the literature.'' \\
{\sc B. D. Ripley}}
\end{flushright}

In this work, we have carried out an extensive test program of a collection
of random number generators, which are commonly used in the applications
of physics. These include public domain programs GGL, RANMAR, RAN3, RCARRY
and R250, a library subroutine G05FAF, and
manufacturer installed routines RAND and RANF.
Also, a sequence of random numbers produced from physical
noise has been included for purposes of comparison.
Our test bench consists of standard statistical tests,
bit level tests and qualitative visual tests.
If we use the first two quantitative tests as criteria,
three of the generators, namely GGL, G05FAF, and R250 display an overall best
performance in all tests, and could thus be recommended for most
applications. They fail statistical tests only once, and produce 31
``good'' bits. Other generators show somewhat less convincing performance in
one or more test category, although RANF performs very well in statistical
tests. If the least significant bits are not important
for the application, both RANF and RANMAR are good choices.
On the other hand, the clear bit level
correlations of RAN3 and poor statistical properties of RCARRY suggest
problems in these generators \cite{Lucher}.
Finally, RAND suffers from an overall
lackluster performance. In Table \ref{tab:summary} we
show a qualitative summary of the performance of all of the generators
in statistical and bit level tests.

Our results also demonstrate the existence of two fundamental problems which
may plague some random number generators. First, a {\em bad implementation}
of a generator algorithm may cause total corruption of the output,
as we have demonstrated for GGL and RAND.
Second problem concerns the
{\em initialization} of generators such as R250, which require several
seed values. This issue has received relatively little attention
in the past, but our results in Section 4.3 demonstrate that as
a result of a bad initialization, correlations in the seeds of R250
transform into the generated random number sequences. Thus even a
good generator can be corrupted by careless use.

Despite the extensive test program presented here, there may still
exist correlations which may be of significance. To this end,
direct physical, application specific
tests of various generators play an important
role and have been conducted
in some special cases
\cite{Kir81,Kal84,Pau84,Mil86,Fer92,Gra93,Vat93b}.
These tests are of particular importance in Monte Carlo simulations,
where physical systems may be very sensitive to spatial
correlations. In particular, it has recently been suggested that
biased results in Monte Carlo simulations of the Ising model
\cite{Fer92} and self - avoiding random walks \cite{Gra93} result from
yet undetected correlations present in
the GFSR($250,103,\oplus$) algorithm (R250 here).
In both cases,
special simulation algorithms were used. In Ref. \cite{Fer92}
the authors suggest that bit level correlations in the most significant
bits of this generator
are responsible for this. However, our results of Sec. 4.3
do not lend support to this claim, since no discernible correlations exist up
to at least 50 numbers apart. We have in fact recently extended the
bit level tests to check correlations up to about 1000 numbers apart,
but find no correlations for our R250 \cite{Vat93b}.
Results of Ref. \cite{Fer92} thus remain
unexplained at the moment. On the other hand, Ref. \cite{Gra93} claims
to confirm these anomalous correlations for GFSR($250,103,\oplus$), and
finds poor performance also for LF(55,24,+) (our RAN3 is
LF(55,24,$-$)). RAN3 spectacularly fails our bit level tests,
which probably explains results of Ref. \cite{Gra93} for
the lagged Fibonacci generator. However,
concerning GFSR(250,103,$\oplus$)
Ref. \cite{Gra93} goes as far as to reinforce the claim \cite{Mar85}
that ``shift register generators using XOR's are among the worst random
number generators and should never have been used''. Based on
our test results this is a somewhat unfair statement, since R250
{\em when properly implemented and initialized}
certainly performs well enough for many
applications. However, we agree with Ref. \cite{Fer92} on the need of
careful physical tests before a ``good quality'' generator is chosen
for a given application.
To unravel possible anomalous correlations in R250,
a new generation of test methods is
clearly needed since no test carried out here can support this
claim. Work in this direction is currently underway \cite{Vat93b}.


\section{Acknowledgments}

We would like to thank Juha Haataja, Jukka Helin,
Pentti Huuhtanen, Kimmo Kaski, Juhani K\"apyaho,
Petri Laurikainen, Jussi Rahola, Robert Swendsen, and Jukka Vanhala
for useful discussions
and Fred James and Manfred Richter
for correspondence. Veikko Nyfors of Cray Finland Oy,
Greg Astfalk of Convex Computer Corporation
and Julie Gulla of Mathworks Inc. have provided information
of the random number generators used in their products.
Aarno Hauru,
Pekka Kyt\"olaakso, Klaus Lindberg and Juha Ruokolainen have given
valuable technical assistance. The Finnish Center for Scientific Computing
and Tampere University of Technology have generously provided the
computing resources. This research was partially supported by the
Academy of Finland and the Foundation of Tampere University of Technology.

\newpage

E-mail addresses:
{\tt Ilpo.Vattulainen@csc.fi,
Kari.Kankaala@csc.fi, \\
jukkas@ee.tut.fi, and ala@phcu.helsinki.fi}

\newpage

%
%

\oddsidemargin=0.0cm
\evensidemargin=0.0cm

\section {Table captions}

{\bf Table 1}

{\sc Table 1}.
Parameters used in the standard tests. $n$ is the length of the
random number sequence and $N$ is the number of times the test
was repeated for the Kolmogorov - Smirnov test. Other parameters
are described in the text.

{\bf Table 2}

{\sc Table 2}.
Results of the statistical tests. Depicted numbers are the values for
the descriptive levels $\delta^+$ and $\delta^-$ from the
Kolmogorov - Smirnov test variables $K^+$ and $K^-$, and $R_{\delta}$ and
$R_K$ denote average
goodness values, as defined in the text (with tests 1 and 2
excluded). The data for $R$'s comes from the first run only.
A generator was considered to fail the test if
the descriptive level was less than 0.05 or more than 0.95. Single,
double and triple consequtive
failures are indicated by single, double, and bold lines,
respectively.

{\bf Table 3}

{\sc Table 3}.
Results of the spectral test for linear congruential generators.
See text for details.

{\bf Table 4}

{\sc Table 4}.
Results of the bit level $d$-tuple and rank tests.
The bits marked failed have failed the test twice.
See text for details.

{\bf Table 5}

{\sc Table 5}.
Results of $d$-tuple and rank tests for R250
initialized with RAN3.

{\bf Table 6}

{\sc Table 6}.
Absolute speeds of the generators on a Cray X-MP/432 EA and a
Convex C3840. S denotes compiling when only scalar optimization
was allowed and V when also vectorizing was allowed.
The timings are in units of microseconds per random number call.
RANF could only be tested on Cray and RAND on
Convex. RAN3 produced erroneous results on Cray.

{\bf Table 7}

{\sc Table 7.}
A summary of the performance of the tested generators in statistical and bit
level tests. For statistical tests, plus denotes at least one case
of one consequtive failure, zero at least one case of two consequtive failures,
and minus at least one case of three consequtive failures. For bit level
tests, plus denotes an impeccable performance, zero the failure of
some of the least significant bits, and minus the failure of more
significant bits for RAND and RAN3. See text for more details.

\clearpage

\section{Figure Captions}

{\bf Figure 1}

{\sc Fig. 1}. Spatial distribution of 20 000 random number pairs in
two dimensions
on a unit square as generated by GGL $(a),(b)$, RAND $(c),(d)$ and
R250 $(e),(f)$. The second figure in each case has a greatly expanded
scale on the $x$ axis.

{\bf Figure 2}

{\sc Fig. 2}. 31 bit binary representations of random numbers produced by
GGL $(a)$ and R250 $(b)$ on a $124 \times 124$ matrix.

{\bf Figure 3}

{\sc Fig 3}. Binary representations of random numbers produced by
R250 when initialized with GGL in single
precision mode.

\clearpage

\begin{table}[htb]
\normalsize \centering
\begin{tabular}{| l | l | r | r | l l l |}
\hline
\multicolumn{2}{| c |}{Test} & \multicolumn{1}{c |}{$n$} &
\multicolumn{1}{c |}{$N$} & \multicolumn{3}{c |}{Other parameters}\\
\hline\hline
(1) & $\chi^2$  	& $100000$ 	& $10000$ 	& $\nu=256$ & & \\
(2) & $\chi^2$  	& $10000$ 	& $10000$ 	& $\nu=128$ & & \\
(3) & Serial test       & $100000$ 	& $1000$ 	& $d=2$	& $\nu=100$ & \\
(4) & Serial test       & $100000$ 	& $1000$ 	& $d=3$	& $\nu=20$ & \\
(5) & Serial test       & $100000$ 	& $1000$ 	& $d=4$ & $\nu=10$ & \\
(6) & Gap test          & $25000$ 	& $1000$	& $\alpha=0$
			& $\beta=0.05$  & $l=30$ \\
(7) & Gap test          & $25000$ 	& $1000$ 	&$\alpha=0.45$
			& $\beta=0.55$  & $l=30$ \\
(8)  & Gap test         & $25000$       & $1000$	&$\alpha=0.95$
			& $\beta=1$     & $l=30$ \\
(9)  & Maximum of $t$   & $2000$ 	& $1000$ 	& $t=5$ & & \\
(10) & Maximum of $t$   & $2000$ 	& $1000$ 	& $t=3$ & & \\
(11) & Collision test   & $16384$ 	& $1000$ 	& $d=2$	& $s=1024$ & \\
(12) & Collision test   & $16384$ 	& $1000$ 	& $d=4$ &$s=32$ & \\
(13) & Collision test   & $16384$	& $1000$ 	& $d=10$&$s=4$ & \\
(14) & Run test         & $100000$	& $1000$	& $l=6$ & & \\
\hline
\end{tabular}
\caption{}
\label{tab:stand_param}
\end{table}

\clearpage

\begin{table}[htb]
\vspace{22.0cm}
\includegraphics{table8_2.ps}
\caption{}
\label{tab:standard}
\end{table}

\clearpage

\begin{table}[htb]
\normalsize \centering
\begin{tabular}{| l | c | c | c | c | c | c | c | c | c | c |}
\hline
 & \multicolumn{2}{c |}{RAND} & \multicolumn{2}{c |}{GGL} &
   \multicolumn{2}{c |}{G05FAF} & \multicolumn{2}{c |}{RANF 1} &
   \multicolumn{2}{c |}{RANF 2} \\
\cline{2-11}
$d$ & $\kappa_{d}$ & $\lambda_{d}$ & $\kappa_{d}$ & $\lambda_{d}$ &
    $\kappa_{d}$ & $\lambda_{d}$ & $\kappa_{d}$ & $\lambda_{d}$ &
    $\kappa_{d}$ & $\lambda_{d}$ \\ \hline\hline
2 & 0.9250 & 15.991 & 0.3375 & 14.037 & 0.8423 & 29.356 & 0.8269 &
						22.8295 & 0.6499 & 22.482 \\
3 & 0.7890 & 10.492 & 0.4412 & 9.319 & 0.7640 & 19.533 & 0.7416 &
						15.069 & 0.7705 & 15.124 \\
4 & 0.7548 & 7.844 & 0.5752 & 7.202 & 0.8472 & 14.260 &  0.3983 &
						10.422 & 0.7071 & 11.250 \\
5 & 0.8041 & 6.386 & 0.7361 & 6.058 & 0.7838 & 11.348 & 0.7307 &
						9.047 & 0.3983 & 8.172 \\
6 & 0.2990 & 3.959 & 0.6454 & 4.903 & 0.6333 & 9.209 & 0.6177 &
						7.339 & 0.6282 & 7.364 \\
7 & 0.4075 & 3.705 & 0.5711 & 4.049 & 0.5540 & 8.382 & 0.6670 &
						6.416 & 0.2375 & 4.926 \\
8 & 0.5762 & 3.705 & 0.6096 & 3.661 & 0.6597 & 7.271 & 0.5642 &
						5.424 & 0.2135 & 4.022 \\\hline
\end{tabular}
\caption{}
\label{tab:spectral}
\normalsize \end{table}

\clearpage

\begin{table}[htb]
\normalsize \centering
\begin{tabular}{| l | l | l | l |}
\hline
Random 		& Failing bits		& Failing bits & Comments of \\
number 		& in the 		& in the rank  & implementation \\
generator 	& {\it d}-tuple test 	& test         & and initialization \\
\hline\hline
GGL 	& none 		& none 		& double precision mode\\
        &		&		& (return integers)\\
RAND 	& 13-31		& 18-31 	& real mode\\
RANF 	& 29-45		& 24,31-45 	& real mode\\
G05FAF	& none		& none 		& double precision mode\\
R250 	& none 		& none 		& integer mode, initialized\\
        &		&		& with GGL in double precision\\
RAN3 	& 1-5,25-30	& 1-5,26-30 	& integer mode\\
RANMAR 	& 25-31		& 25-31		& real mode\\
RCARRY 	& 25-31		& 25-31		& real mode\\
PURAN II & none		& none		& integer mode\\
\hline
\end{tabular}
\caption{}
\label{tab:bitcorr1}
\end{table}

\clearpage

\begin{table}[htb]
\normalsize \centering
\begin{tabular}{|l|l|l|}
\hline
Random		& Failing bits 		& Failing bits \\
number 		& in the 		& in the \\
generator 	& {\it d}-tuple test 	& rank test \\
\hline\hline
R250 & 1 - 2, 27 - 31 & 1, 27 - 31 \\
RAN3 & 1 - 5, 25 - 30 & 1 - 5, 26 - 30 \\
\hline
\end{tabular}
\caption{}
\label{tab:bitcorr2}
\end{table}

\clearpage

\begin{table}[htb]
\normalsize \centering
\begin{tabular}{| l | c | c | c | c | c |}
\hline
Generator & Optimization & \multicolumn{2}{c |}{Cray} &
                               \multicolumn{2}{c |}{Convex} \\ \cline{3-6}
     & & $n=1$ & $n=1000$ & $n=1$ & $n=1000$ \\
\hline\hline
GGL    &  S  & 2.218 & 2.731 & 4.420 & 2.379 \\
       &  V  & 2.465 & 2.029 & 5.676 & 2.381 \\ \hline
RAND   &  S  & $-$   & $-$   & 4.446 & 4.582 \\
       &  V  & $-$   & $-$   & 6.661 & 4.369 \\ \hline
RANF   &  S  & 1.466 & 1.582 & $-$   & $-$   \\
       &  V  & 1.536 & 0.020 & $-$   & $-$   \\ \hline
G05FAF &  S  & 4.556 & 0.422 & 4.384 & 0.571 \\
       &  V  & 4.442 & 0.365 & 6.321 & 0.559 \\ \hline
R250   &  S  & 260.0 & 1.672 & 126.7 & 1.094 \\
       &  V  & 10.88 & 0.055 & 55.87 & 0.476 \\ \hline
RAN3   &  S  & 4.711 & $-$   & 3.987 & 2.177 \\
       &  V  & 3.563 & $-$   & 4.881 & 1.608 \\ \hline
RANMAR &  S  & 7.132 & 3.407 & 5.672 & 1.932 \\
       &  V  & 4.801 & 1.053 & 5.742 & 1.508 \\ \hline
RCARRY &  S  & 6.486 & 2.455 & 4.956 & 1.211 \\
       &  V  & 3.962 & 0.728 & 4.537 & 0.899 \\ \hline
\end{tabular}
\caption{}
\label{tab:speed}
\normalsize \end{table}

\clearpage

\begin{table}[htb]
\normalsize \centering
\begin{tabular}{| l | c | c | c | c | c | c | c | c |}
\hline
\multicolumn{1}{| l |}{Test} & \multicolumn{8}{c |}{Random number generator} \\
\cline{2-9}
method 		& GGL & RAND & RANF & G05FAF & R250 & RAN3 & RANMAR & RCARRY \\
\hline\hline
Standard tests	& $+$ & $0$ & $+$ & $+$ & $+$ & $0$ & $-$ & $-$ \\
Bit level tests	& $+$ & $-$ & $0$ & $+$ & $+$ & $-$ & $0$ & $0$ \\
\hline
\end{tabular}
\caption{}
\label{tab:summary}
\end{table}

\end{document}